\documentclass[12pt,a4paper,english]{article}
\usepackage[T1]{fontenc}
\usepackage[latin2]{inputenc}
\usepackage[english]{babel}
\usepackage{amsmath}
\usepackage{amsfonts}
\usepackage{indentfirst}
\usepackage[dvips]{graphicx}

\begin{document}

\sloppy


\begin{flushright}
\begin{tabular}{l}
hep-th/0606055  \\
IFT-UW-2006-10 \\
ITFA-2006-21
\\ [.3in]
\end{tabular}
\end{flushright}

\begin{center}
\Large{ \bf Crystal Model for the Closed Topological Vertex Geometry}
\end{center}

\begin{center}

\bigskip

Piotr Su\l kowski

\bigskip

\emph{Institute for Theoretical Physics, Warsaw University} \\
\emph{ul. Ho\.za 69, PL-00-681 Warsaw, Poland} \\

\medskip

\emph{and}

\medskip

\emph{Institute for Theoretical Physics, University of Amsterdam} \\
\emph{Valckenierstraat 65, 1018 XE Amsterdam, The Netherlands} \\

\bigskip

\centerline{
\emph{Piotr.Sulkowski@fuw.edu.pl} }

\smallskip
 \vskip .3in \centerline{\bf Abstract}
\smallskip

\end{center}

The topological string partition function for the neighbourhood of three spheres meeting at one point in a Calabi-Yau threefold, the so-called 'closed topological vertex', is shown to be reproduced by a simple Calabi-Yau crystal model which counts plane partitions inside a cube of finite size. The model is derived from the topological vertex formalism. This derivation can be understood as 'moving off the strip' in the terminology of hep-th/0410174, and  offers a possibility to simplify topological vertex techniques to a broader class of Calabi-Yau geometries. To support this claim a flop transition of the closed topological vertex is considered and the partition function of the resulting geometry is computed in agreement with general expectations.


\newpage

\section{Introduction}

Topological string amplitudes can be presented in a variety of ways depending on their interpretation and the physical picture one is interested in. Their basic definition in terms of Gromov-Witten invariants relates to the worldsheet description and has been given precise mathematical foundations \cite{mirror}. The other reformulations are based on conjectural physical dualities and refer to a target space point of view. M-theory interpretation reveals their integrality properties encoded in the Gopakumar-Vafa invariants related to the counting of BPS states \cite{G-V}. A connection to Chern-Simons theory - whose solution is well known - via the open-closed duality \cite{G-V-transition,marcos} allows to construct the topological vertex \cite{vertex,B-vertex,math-vertex} from which solutions of closed topological strings on arbitrary toric Calabi-Yau manifolds can be built. Yet another line of development relates topological amplitudes to the Donaldson-Thomas invariants, Calabi-Yau crystals and their quantum foam interpretation \cite{ok-re-va,foam,mnop,va-sa,ani,okuda,ps}.

The object we focus on in this note is the so-called closed topological vertex $\mathcal{C}$ --- a local Calabi-Yau neighbourhood of three $\mathbb{P}^1$'s meeting in one point. This is an example of a nontrivial but exactly solvable geometry, for which the above mentioned pictures can be given explicitly and some of them are already known \cite{cremona,GW-curves}. The main contribution we provide is a crystal model for the closed topological vertex, which extends the class of known Calabi-Yau crystals. This new model naturally reproduces the Gopakumar-Vafa invariants, which is a general property observed before. We prove that the model introduced indeed corresponds to the closed topological vertex $\mathcal{C}$ by relating it explicitly to the topological vertex computations.

Apparently, $\mathcal{C}$ belongs to the class of 'off-strip' geometries, whose dual toric diagrams cannot be presented as a triangulation of a rectangle. On the other hand, the topological vertex calculations for a special class of geometries which can be presented on a strip have been vastly simplified in \cite{strip}. Thus the method we use offers a possibility of simplifying those rules for a broader class of geometries. We provide another example of an 'off-strip' geometry by considering a flop transition of $\mathcal{C}$. The partition function for the resulting geometry $\mathcal{C}^{flop}$ can also be computed in a way parallel to the computation for $\mathcal{C}$ before the flop, and it is possible to determine both Gopakumar-Vafa invariants and classical contributions for $\mathcal{C}^{flop}$ in terms of those for $\mathcal{C}$.

The plan of the paper is as follows. In section \ref{sec-closed-ver} the geometry of the closed topological vertex is introduced, and its topological string partition function is presented and discussed from various points of view. Section \ref{sec-crystals} starts with a brief presentation of the known Calabi-Yau crystal models which is followed by an introduction of a new model. The properties of this new model are discussed, and the connection with the closed topological vertex is explained. In section \ref{s-vertex} the model is explicitly related to the topological vertex calculation and its solution as an 'off-strip' geometry is presented; a possibility of generalization to other 'off-strip' geometries is accompanied by a careful analysis of a flop transition of $\mathcal{C}$. Section \ref{sec-discuss} contains the discussion of the results and directions for further studies. The relevant notation and the properties of Schur functions we need are assembled in appendix \ref{app-schur}. Finally appendix \ref{app-strip} summarizes the rules for computations on a strip. 


\section{The closed topological vertex geometry} \label{sec-closed-ver}

\begin{figure}[htb]
\begin{center}
\includegraphics[width=0.4\textwidth]{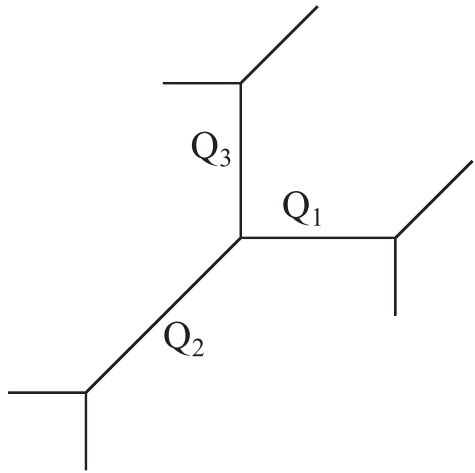}
\caption{The closed topological vertex $\mathcal{C}$.} \label{fig-closed-ver}
\end{center}
\end{figure}

The closed topological vertex $\mathcal{C}$ is a toric Calabi-Yau threefold, consisting of three $\mathbb{P}^1$'s meeting in one point, with local neighbourhood of each sphere being isomorphic to a sum of line bundles $\mathcal{O}(-1)\oplus \mathcal{O}(-1)$. This geometry has been discussed in \cite{cremona,GW-curves}, and can be understood as a particular resolution of $\mathbb{C}^3/\mathbb{Z}_2\times\mathbb{Z}_2$ singularity \cite{branes-resolve,F-theory}. It is convenient to introduce quantities $Q_i=e^{-t_i}$ related to K\"ahler parameters $t_i$, $i=1,2,3$ which correspond to sizes of the spheres. The toric diagram of the closed topological vertex is shown in figure \ref{fig-closed-ver}. 

The main object of considerations below is the partition function of $\mathcal{C}$. There is a variety of ways how to compute or present topological string partition functions, and in the case of the closed topological vertex it is not difficult to make them explicit. Some of these results are already done in literature, so first we briefly quote them to set later computations in a proper context.

Let us recall the A-model topological string partition function for arbitrary Calabi-Yau manifold $X$ is of the form
\begin{eqnarray}
Z^X_{top} & = & e^{F^X_{top}} = M(q)^{\chi(X)} e^{F^X_{class}+F^X}, \label{Z-top} \\
F^{X}_{class} & = & \sum\frac{1}{6g_s^2} a_{ijk}\,t_i t_j t_k + \sum \frac{1}{24} b_i t_i, \label{F-class} \\
F^X & = & \sum_{g\geq 0} \sum_{\beta\neq 0} g_s^{2g-2}\, N^{g}_{\beta}\, Q^{\beta}, \label{grom-wit}
\end{eqnarray}
with the notation as follows. $M(q)$ is McMahon function given in (\ref{crystal-C3}) with $q=e^{-g_s}$, and $\chi(X)$ is Euler characteristic of $X$. $Q_i=e^{-t_i}$ are related to K\"ahler parameters of the manifold $t_i$. The free energy $F^X_{top}$ is a sum of $F^X_{class}$ with polynomial dependence on $t_i$ (in particular with genus zero cubic terms encoding  classical intersection numbers $a_{ijk}$ and genus one terms related to $c_2(X)$) and contributions from worldsheet instantons $F^X$. The latter are encoded in terms of Gromov-Witten invariants $N^{g}_{\beta}$, which roughly  count maps from a genus $g$ Riemann surface into a curve in a Calabi-Yau manifold of a certain class $\beta= (d_i) \in H_2(X,\mathbb{Z})$, and $Q^{\beta}=\prod Q_i^{d_i}$. In most of what follows we restrict our attention to instanton contributions $Z^{X} = \exp F^{X}$ as they arise naturally in crystal models.

For the closed topological vertex $\beta=(d_1,d_2,d_3)$, with degree $d_i$ corresponding to the $i$'th sphere, and (\ref{grom-wit}) reads
$$
F^{\mathcal{C}} = \sum_{g\geq 0} \sum_{d_1,d_2,d_3} g_s^{2g-2}\, N^{g}_{d_1,d_2,d_3}\, Q_1^{d_1} Q_2^{d_2} Q_3^{d_3}. \label{grom-wit-closed}
$$
These local Gromov-Witten invariants (local for $\mathcal{C}$ being noncompact) can be rigorously derived via the Cremona transform by identification with known invariants of a relevant blow-up of a projective space. This approach was originally introduced in this context in \cite{cremona-K3}, and applied to the closed topological vertex in \cite{cremona} with the result
\begin{eqnarray}
N^{g}_{d,0,0} & = & N^{g}_{0,d,0} = N^{g}_{0,0,d} = N^{g}_{d,d,d} = -N^{g}_{d,d,0}  = -N^{g}_{d,0,d} = -N^{g}_{0,d,d} = \nonumber \\
& = & d^{2g-3} \frac{|B_{2g}|}{2g\,(2g-2)!}, \qquad \textrm{for}\ d\neq 0, \label{GW-P1}
\end{eqnarray}
and all other $N^{g}_{d_1,d_2,d_3}$ vanishing. Apparently, $N^{g}_{d,0,0}=N^g_d$ is equal to the invariant $N^g_d$ for a super-rigid $\mathbb{P}^1$ (i.e. resolved conifold) derived in \cite{localP1}. 

The structure of the topological string partition function is conjecturally encoded in integer Gopakumar-Vafa invariants \cite{G-V} $n^g_{\beta}$, so that (\ref{grom-wit}) takes the form
$$
F^{X} = \sum_{g\geq 0} \sum_{\beta=(\beta_i)} \sum_{n\geq 1} n^g_{\beta} \frac{-Q^{n\beta}}{n\,[n]^{2-2g}} 
$$
with $[n]=q^{n/2}-q^{-n/2}$. This nice integrality property is balanced by the fact that there is no general recipe to compute them, which is not unrelated to difficulties in their formal mathematical definition. It is instructive to consider first the resolved conifold geometry with a single K\"ahler parameter $Q=e^{-t}$ corresponding to the size of $\mathbb{P}^1$. In this case there is only one non-trivial Gopakumar-Vafa invariant $n^0_1=1$ which encodes the information of all Gromov-Witten invariants $N^g_d = N^{g}_{d,0,0}$ given in (\ref{GW-P1}), and the free energy can be rewritten as
\begin{equation}
F^{conifold} = -\sum_{g\geq 0} g_s^{2g-2} \frac{|B_{2g}|}{2g\,(2g-2)!} Li_{3-2g}(Q) = \sum_{n\geq 1}\frac{-Q^{n}}{n [n]^{2}}, \label{F-conifold}
\end{equation}
where in genus expansion a polylogarithm arises $Li_k(Q)=\sum_{n\geq 1} \frac{Q^n}{n^k}$.

The structure of the closed topological vertex partition function is similar to $F^{conifold}$, in a sense that there is also only a finite number of non-vanishing Gopakumar-Vafa invariants
$$
n^0_{1,0,0}= n^0_{0,1,0} = n^0_{0,0,1} =  - n^0_{1,1,0} = - n^0_{1,0,1} = - n^0_{0,1,1} = n^0_{1,1,1} = 1
$$
and they correspond respectively to single spheres, each possible pair of them and the entire triple. Thus the instanton part of the partition function for $\mathcal{C}$ can be written as
\begin{equation}
Z^{\mathcal{C}} = \exp\sum_{n>0}\frac{-Q_1^n-Q_2^n-Q_3^n + Q_1^n Q_2^n + Q_1^n Q_3^n + Q_2^n Q_3^n - Q_1^n Q_2^n Q_3^n}{n[n]^2}. \label{closed-ver}
\end{equation}
Of course this result is consistent with (\ref{GW-P1}). This can also be written as a product formula, as described in general in \cite{HIV}. In fact, it turns out that the crystal model we introduce in the next section naturally computes the result in the form related both to the above Gopakumar-Vafa expansion and a product formula, which is a fact already stressed in \cite{ps}. 

There is a very effective way to compute topological string quantities for toric threefolds in terms of the topological vertex formalism \cite{vertex,B-vertex}. This is based on a conjectural geometric transition to open topological strings and their relation to Chern-Simons theory \cite{marcos}. There is also an accompanying mathematical formulation \cite{math-vertex}. $Z^{\mathcal{C}}$ has been computed in these 'physical' and 'mathematical' formalisms in \cite{GW-curves}, and shown to agree with (\ref{closed-ver}). In section \ref{s-vertex} we will also compute $Z^{\mathcal{C}}$ using 'physical' topological vertex, but in a way which is simpler and faster than it is done in \cite{GW-curves}. In fact the main motivation behind the calculation we present is it makes an immediate connection with the crystal model which we present next. And then - last but not least - the method presented here suggests how to generalize formalism of \cite{strip} to 'off-strip' geometries.

So far we focused only on the instanton contributions $F^{\mathcal{C}}$ which have already been derived in literature. The classical part $F_{class}^{\mathcal{C}}$ will be discussed in section \ref{subsec-flop} together with the analysis of the flop transition.


\section{Crystal models} \label{sec-crystals}

\subsection{The idea}

The notion of Calabi-Yau crystals appeared first in \cite{ok-re-va}. It was based on the observation that the topological string partition function for $\mathbb{C}^3$ is equal to a generating function $Z$ for a classical system which consists of 3-dimensional partitions which can fill a positive octant of $\mathbb{R}^3$. Allowed 3-dimensional configurations of partitions are weighted by a number of boxes they are built of,
\begin{equation}
Z = \sum_{\pi - 3d\, partition} q^{|\pi|}. \label{generating}
\end{equation}
This partition function can be computed in a simple way in a transfer matrix formalism, carefully presented in \cite{ok-re-va}. Essentially, it amounts to slicing the $\mathbb{R}^3$ octant by $x=y$ planes. Then each 3-dimensional partition turns into a sequence of 2-dimensional partitions, the neighbouring ones necessarily satisfying an interlacing relation $\succ$ which is a consequence of the allowed 3-dimensional configurations. One can encode each 2-dimensional partition as a state of a Fermi sea $|\mu \rangle$ in a standard way, and then use the operators 
\begin{equation}
\Gamma_{\pm}(z)=\exp \sum_{n>0} \frac{z^{\pm n}}{n}\alpha_{\pm n} \label{gamma}
\end{equation}
satisfying $\Gamma_{-}(1) |\mu \rangle = \sum_{\nu \succ \mu} |\nu\rangle$ to compute
\begin{eqnarray}
Z^{\mathbb{C}^3}(q) & = & \langle 0|\prod_{n>0}\Gamma_{+}(q^{n-1/2})\,\prod_{m>0}\Gamma_{-}(q^{-(m-1/2)})|0\rangle = \nonumber \\
& = & M(q) = \prod_{n=1}^{\infty} \frac{1}{(1-q^n)^n} = \exp{\sum_{n>0} \frac{1}{n[n]^2}}, \label{crystal-C3}
\end{eqnarray}
where $M(q)$ is McMahon function. The computation is straightforward and amounts to using commutation relations
\begin{equation}
\Gamma_{+}(z)\Gamma_{-}(z')=\frac{1}{1-z/z'}\Gamma_{-}(z')\Gamma_{+}(z). \label{commute}
\end{equation}

The appearance of such a statistical model in connection with topological strings has found two explanations. On one hand it was interpreted as gravitational quantum foam \cite{foam} and expressed in terms of 6-dimensional gauge theory. On the other, it was shown in \cite{okuda} that the Chern-Simons partition function on $S^3$ for $U(N)$ gauge group can be essentially rewritten in terms of a crystal model, and then the relation to closed topological strings arises from 't Hooft duality. The result (\ref{crystal-C3}) arises as  $N\to \infty$ limit of the model in \cite{okuda}, which is valid for arbitrary $N$. For any finite $N$ the necessary condition is to count only these 3-dimensional diagrams which have at most $N$ boxes in one direction (and no restriction in two other directions). In other words, we cut-off the positive octant where 3-dimensional partitions were considered by a 'wall' at position $x=N$. In the transfer matrix formalism, the introduction of such a wall is possible in two ways; in terminology of \cite{okuda} either by considering closed-string slicing or open-string slicing. The former is obtained by inserting only $N$ operators in one direction in the expression (\ref{crystal-C3}); the letter amounts to inserting a projector operator $\mathbf{1_{d^t\leq N}}$ between the sequence of $\Gamma_+$ and $\Gamma_-$ operators
\begin{eqnarray}
Z^{\mathbb{P}^1} & = & \langle 0|\prod_{n>0}\Gamma_{+}(q^{n-1/2})\,\prod_{m=1}^{N}\Gamma_{-}(q^{-(m-1/2)})|0\rangle = \nonumber \\
& = & \langle 0|\prod_{n>0}\Gamma_{+}(q^{n-1/2})\,\mathbf{1_{d^t\leq N}}\, \prod_{m>0}\Gamma_{-}(q^{-(m-1/2)})|0\rangle = \nonumber \\
& = & M(q) \exp\sum_{n>0}\frac{-Q^n}{n[n]^2}, \label{crystal-P1}
\end{eqnarray}
where $Q=q^N$. The final result is the closed string partition function for the resolved conifold $Z^{\mathbb{P}^1}$, as it should. The operator $\mathbf{1_{d^t\leq N}}$ can be written in terms of some coherent states and integration over $U(N)$, but as we don't need its explicit form we refer the interested reader to \cite{okuda}.

\begin{figure}[htb]
\begin{center}
\includegraphics[width=0.8\textwidth]{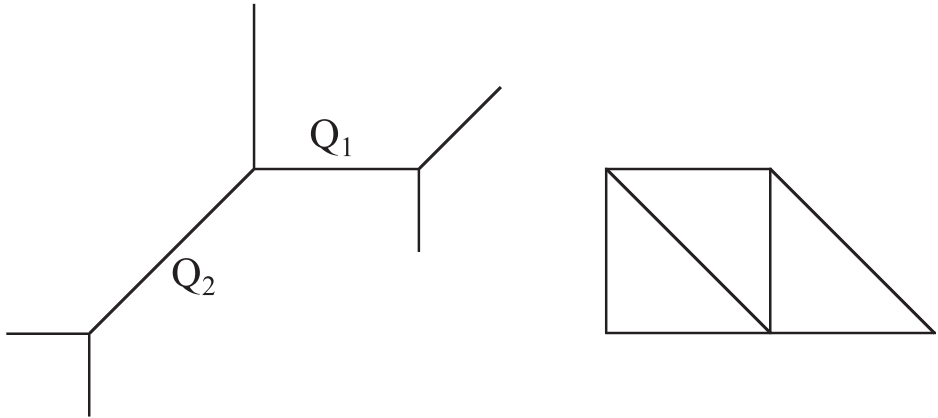}
\caption{Toric diagram (left) and its dual (right) for the double-$\mathbb{P}^1$.} \label{fig-double-P1}
\end{center}
\end{figure}

We should stress that this is the first line of (\ref{crystal-P1}) which can be explicitly evaluated due to (\ref{commute}) relations. It has an obvious generalization, which amounts to inserting finite numbers $N_1,N_2$ of \emph{both} $\Gamma_{\pm}$ operators, and corresponds to 3-dimensional partitions restricted to the 'well' of rectangular size $N_1\times N_2$ and infinite hight. Explicit computation in \cite{ps} using topological vertex formalism shows a generating function for such a classical system is equal to a partition function for a toric threefold called 'double-$\mathbb{P}^1$' shown in figure \ref{fig-double-P1}, which is a local neighbourhood of two $\mathbb{P}^1$'s meeting in a single point, with K\"ahler parameters $t_{1,2}$ determining $\mathbb{P}^1$'s sizes given by $Q_{1,2} = e^{-t_{1,2}} = q^{N_{1,2}}$
\begin{eqnarray}
Z^{double-\mathbb{P}^1} & = & \langle 0|\prod_{n=1}^{N_2}\Gamma_{+}(q^{n-1/2})\,\prod_{m=1}^{N_1}\Gamma_{-}(q^{-(m-1/2)})|0\rangle = \nonumber \\
& = & M(q) \exp\sum_{n>0}\frac{-Q_1^n -Q_2^n + Q_1^n Q_2^n}{n[n]^2}. \label{crystal-P1P1}
\end{eqnarray}
This can also be immediately obtained from (\ref{Zalpha}) with $\alpha=\bullet$, in which case the products over $k$ in (\ref{amir-bracket}) are trivial and we are left with the same exponential factors as above. We now wish to pursue this argument further.

\subsection{The cube model}

\begin{figure}[htb]
\begin{center}
\includegraphics[width=0.5\textwidth]{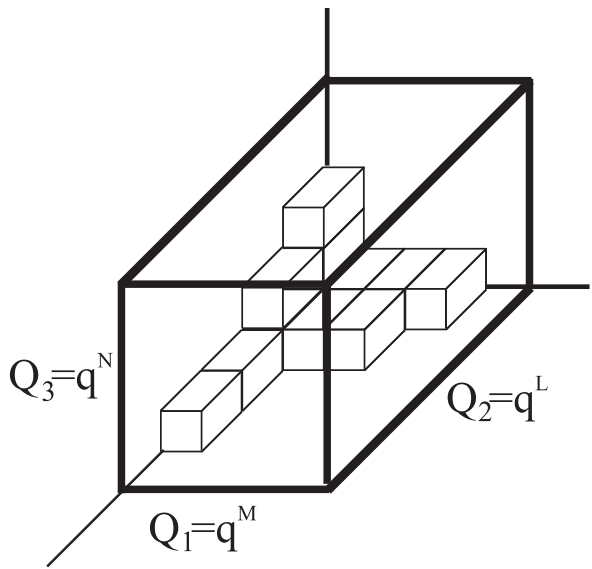}
\caption{The cube crystal model of finite size  $M \times L \times N$.} \label{fig-cube}
\end{center}
\end{figure}

The model we wish to introduce is a natural continuation of the above presentation. We consider all 3-dimensional partitions which fit into a finite cube of size $M \times L \times N$, and ask what is the corresponding generating function $Z^{cube}$, given by (\ref{generating}) with the present restriction on $\pi$. Let us remark there is only a finite number of terms in $Z^{cube}=1+\ldots + q^{LMN}$, the last one corresponding to the highest power of $q$. In other words, we introduce three 'walls' at positions $x=M$, $y=L$, $z=N$, as illustrated in figure \ref{fig-cube}.

In this case the partition function can also be written in the transfer matrix formalism, with two walls represented by finite number of $\Gamma_{\pm}$ and the third wall by the projection $Z^{\mathbb{P}^1}$
\begin{equation}
Z^{cube} = \langle 0|\prod_{n=1}^{L}\Gamma_{+}(q^{n-1/2})\,\mathbf{1_{d^t\leq N}}\, \prod_{m=1}^{M}\Gamma_{-}(q^{-(m-1/2)})|0\rangle. \label{crystal-cube}
\end{equation}

Our claim is this generating function is equal to the closed topological vertex partition function (\ref{closed-ver}) up to the McMahon function
\begin{equation}
Z^{cube} = M(q) Z^{\mathcal{C}}, \label{claim}
\end{equation}
with identification of parameters
\begin{equation}
Q_1 = q^M,\qquad Q_2 = q^L, \qquad Q_3 = q^N.  \label{Q123-MLN}
\end{equation}

The explicit evaluation of the correlator (\ref{crystal-cube}) is nontrivial, due to a complicated form of the projector $\mathbf{1_{d^t\leq N}}$. Fortunately, the generating function for plane partitions in a finite box has been well known since originally derived by combinatorial methods by McMahon \cite{mcmahon}; another combinatorial proof is presented in \cite{macdonald}. Thus we can just use this result, which reads:
\begin{equation}
Z^{cube} =  Z_1 Z_2 = \prod_{(i,j)\in M^L} \frac{1-q^{N+i+j-1}}{1-q^{h(i,j)}}, \label{Z1Z2} 
\end{equation}
where for later convenience we introduce two factors
\begin{eqnarray}
Z_1 & = & \prod_{(i,j)\in M^L} \frac{1}{1-q^{h(i,j)}}, \nonumber \\
Z_2 & = & \prod_{(i,j)\in M^L} (1-q^{N+i+j-1}), \nonumber 
\end{eqnarray}
and $M^L$ denotes 2-dimensional partition with L rows of the same length $M$, which is a base of the cube, and $h(i,j)$ is the hook-length of its $(i,j)$ element (\ref{hook}). Using the elementary series
$$
\log(1-a)=-\sum_{k>0}\frac{a^{k}}{k}
$$
to rewrite products as exponentials we arrive at the following form of the above factors
\begin{eqnarray}
Z_1 & = & \exp\sum_{n>0}\frac{1-Q_1^n -Q_2^n + Q_1^n Q_2^n}{n[n]^2}, \label{Z1-exp}\\
Z_2 & = & \exp\sum_{n>0}\frac{-Q_3^n + Q_1^n Q_3^n + Q_2^n Q_3^n - Q_1^n Q_2^n Q_3^n}{n[n]^2}. \label{Z2-exp}
\end{eqnarray}
Remarkably, the product (\ref{Z1Z2}) of the above two factors indeed reproduces the closed topological vertex partition function (\ref{closed-ver}) up to the McMahon function, as claimed in (\ref{claim}). In the next section we derive this result from the topological vertex point of view, in a way which makes relation to the crystal result (\ref{Z1Z2}) explicit.


\section{Topological vertex and 'off-strip' geometries} \label{s-vertex}

\subsection{Derivation}

Now we derive the result (\ref{closed-ver}) from the topological vertex formalism, in a way which gives this result in the form explicitly related to (\ref{Z1Z2}). In computation we use conventions and various identities given in appendix \ref{app-schur}, as well as a variety of well-known properties of Schur functions, assembled e.g. in appendices in reference \cite{ps}. In fact the result can be obtained much faster, if certain sums in the amplitude are automatically performed using additional machinery of \cite{strip}, which we also review in appendix \ref{app-strip}. It turns out the derivation presented below can be understood as an extension of that machinery to a more general situation. Nonetheless, for completeness we first derive the closed topological vertex partition function from first principles. 

The basic topological vertex amplitude is
\begin{equation}
C_{R_1 R_2 R_3} = q^{\frac{1}{2}(\kappa_{R_2}+\kappa_{R_3})} s_{R_{2}^{t}}(q^{\rho})
\, \sum_{P} s_{R_{1}/P}(q^{ R_{2}^{t}+\rho}) s_{R_{3}^{t}/ P}(q^{ R_{2}+\rho}).
\label{vertex}
\end{equation}
It will turn out convenient to use cyclic symmetry to write the amplitude for the closed topological vertex from figure \ref{fig-closed-ver} as
\begin{eqnarray}
Z^{\mathcal{C}} & = & \sum_{R_1,R_2,R_3} C_{R_2 R_3 R_1} C_{R_1^t\bullet\bullet} C_{R_2^t\bullet\bullet} C_{R_3^t\bullet\bullet} (-Q_1)^{|R_1|}  (-Q_2)^{|R_2|} (-Q_3)^{|R_3|}  \nonumber \\
& & \qquad \qquad s_{R_1}(-Q_1 q^{\rho}) s_{R_2^t}(-Q_2 q^{\rho}) s_{R_3}(-Q_3 q^{\rho}) = \nonumber \\
& = & \sum_{R_1,R_2,R_3,P} \Big[s_{P}(-Q_1 q^{\rho}) s_{P}(-Q_2 q^{\rho})\Big]\Big[ s_{R_3}(q^{\rho}) s_{R_3^t}(-Q_3 q^{\rho})  \Big] \nonumber \\ 
& & \qquad \qquad \Big[ s_{R_2}(q^{R_3^t+\rho}) s_{R_2^t}(-Q_2 q^{\rho}) \Big] \Big[s_{R_1^t}(q^{R_3+\rho}) s_{R_1}(-Q_1 q^{\rho}) \Big].  \nonumber
\end{eqnarray}


This can be rewritten using (\ref{sum-schur-RRbis}) in the form
$$
Z^{\mathcal{C}} = \exp \Big[ \sum_n \frac{-Q_1^k -Q_2^k + Q_1^k Q_2^k}{n[n]^2} \Big] \times
$$
\begin{equation}
\times \sum_{R_3} s_{R_3}(q^{\rho}) s_{R_3^t}(q^{\rho}) (-Q_3)^{|R_3|} \prod_k (1-Q_1 q^k)^{C_k(R_3)} (1-Q_2 q^k)^{C_k(R_3^t)}. \label{cl-vert-compute-2}
\end{equation}
The exponent factor which arises above is equal to $Z_1$ (\ref{Z1-exp}), which is double-$\mathbb{P}^1$ partition function (up to the McMahon function). Thus we have to show the sum over $R_3$ above reproduces $Z_2$ (\ref{Z2-exp}). We use (\ref{schur-hooks})  to rewrite (\ref{cl-vert-compute-2}) as 
\begin{equation}
Z^{\mathcal{C}} = \frac{Z_1}{M(q)} \sum_{R_3} (-Q_3)^{|R_3|} X_{R_3} Y_{R_3},
\end{equation}
where
\begin{eqnarray}
X_{R_3} & = & s_{R_3^t}(q^{\rho}) \prod_k (1-Q_1 q^k)^{C_k(R_3)} =  \nonumber \\
& = & (-1)^{|R_3|} q^{|R_3|/2+n(R_3)} \prod_{(i,j)\in R_3} \frac{1-q^{M+j-i}}{1-q^{h(i,j)}}
\end{eqnarray}
and similarly
$$
Y_{R_3} = (-1)^{|R_3|} q^{|R_3^t|/2+n(R_3^t)} \prod_{(i,j)\in R_3^t} \frac{1-q^{L+j-i}}{1-q^{h(i,j)}}
$$
where we used identification (\ref{Q123-MLN}).

Finally, the crucial step is to rewrite $X_{R_3}$ and $Y_{R_3}$ using the identity (\ref{schur-finite}) for a Schur function with finite number of variables
\begin{eqnarray}
Z^{\mathcal{C}} & = & \frac{Z_1}{M(q)} \sum_{R_3} (-q Q_3)^{|R_3|} s_{R_3}(1,q,q^2,\ldots,q^{M-1}) s_{R_3^t}(1,q,q^2,\ldots,q^{L-1}) \nonumber \\
& = & \frac{Z_1}{M(q)} \prod_{i=1,\ldots L;\, j=1, \ldots, M} (1-Q_3 q^{j+i-1}) = \frac{Z_1 Z_2}{M(q)} \label{cl-vert-compute-3}
\end{eqnarray}
where the sum of Schur functions (\ref{schur-sum}) was used in the last line. Because $Z_1 Z_2 = Z^{cube}$, we indeed obtain (\ref{claim})
$$
Z^{cube} = M(q) Z^{\mathcal{C}}.
$$

In the above derivation the same factors as in the crystal model arise, i.e. $Z_1$ associated with double-$\mathbb{P}^1$ partition function and $Z_2$ 'implementing' the third $\mathbb{P}^1$. In this form it also becomes a trivial observation that in $N \to \infty$ limit the proper result for double-$\mathbb{P}^1$ is recovered (\ref{crystal-P1P1}). 

Let us also remark on McMahon $M(q)$ factors. It is known the topological vertex computations do not produce this contribution, so to obtain the full topological string partition function (\ref{Z-top}) one should introduce one factor of $M(q)$ for each topological vertex $C_{PQR}$ used in the vertex computation of the amplitude \cite{foam}. On the other hand, in our crystal model just a single factor of $M(q)$ arises as the plane partitions it counts are anchored in one particular corner of the cube. This is why there is one factor of $M(q)$ in (\ref{claim}). 


\subsection{Moving off the strip}

It is interesting to relate this result to the formalism developed in \cite{strip} which allows to perform computations for toric geometries whose dual diagrams are given by a triangulation of a rectangle (or a 'strip' - hence the terminology). We briefly review this machinery in appendix \ref{app-strip}. A simple example of a geometry of this type is the double-$\mathbb{P}^1$ from figure \ref{fig-double-P1}, whose dual diagram is undoubtedly a triangulation of a strip.

\begin{figure}[htb]
\begin{center}
\includegraphics[width=0.4\textwidth]{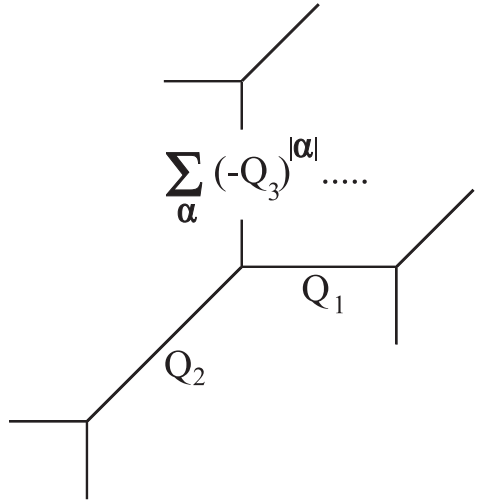}
\caption{The closed topological vertex as a strip with an additional $\mathbb{P}^1$ (of the size determined by $Q_3$) attached.} \label{fig-closed-ver-glue}
\end{center}
\end{figure}

Let us notice that the toric diagram for the closed topological vertex can be understood as a double-$\mathbb{P}^1$ with one additional sphere attached. This is shown in figure \ref{fig-closed-ver-glue}, with double-$\mathbb{P}^1$ given by spheres $Q_1-Q_2$ and the additional sphere denoted by $Q_3$. Even though the full diagram for the closed topological vertex cannot be drawn on a strip, a diagram for the double-$\mathbb{P}^1$ part can as presented above. Thus the partition function $Z^{\mathcal{C}}$ can be written as
$$
Z^{\mathcal{C}} = \sum_{\alpha} s_{\alpha^t}(q^{\rho}) (-Q_3)^{|\alpha|} Z_{\alpha},
$$
where $Z_{\alpha}$ is the factor for a double-$\mathbb{P}^1$ with one nontrivial representation, and is derived in appendix (\ref{Zalpha})
\begin{eqnarray}
Z_{\alpha} & = & s_{\alpha}\, \{\bullet \alpha \}_{Q_1} [\bullet \bullet ]_{Q_1 Q_2} \{\alpha^t \bullet \}_{Q_2} = \nonumber \\
& = & \frac{Z_1}{M(q)} \, s_{\alpha}(q^{\rho}) \prod_k (1-Q_1 q^k)^{C_k(\alpha)} (1-Q_2 q^k)^{C_k(\alpha^t)},
\end{eqnarray}
which altogether reproduces (\ref{cl-vert-compute-2}), and now the calculation continues as above and reproduces (\ref{cl-vert-compute-3}).

So, in this way we managed to 'move off the strip', which technically boils down to performing a sum of Schur functions with finite number of arguments (\ref{cl-vert-compute-3}). It thus seems likely this result might be generalized to a broader class of non-strip-like Calabi-Yau manifolds, and is interesting to pursue further. Below we consider an example of another 'off-strip' geometry related by a flop to the closed vertex, for which a partition function can also be derived using these methods.

\subsection{Flop transition} \label{subsec-flop}

The closed topological vertex consists of three $\mathbb{P}^1$'s with local bundles isomorphic to the resolved conifold. As is well known, such a bundle may undergo a flop transition. For the resolved conifold with K\"ahler parameter $t$ and $Q=e^{-t}$, a flop may be understood as a continuation to negative values of $t$. The partition function is invariant under this process and this should be seen order by order in genus expansion. To get a partition function of the flopped geometry, after the analytic continuation to negative $t$ one should expand the result again in positive powers of $Q$. For the resolved conifold the geometry before and after the transition is the same, which allows to fix the polynomial dependence of the free energy on $t$ \cite{G-V-transition}. In the case of the closed topological vertex the geometries before and after the flop are different, but it is possible to determine classical contribution to the free energy of $\mathcal{C}^{flop}$ in terms of those of $\mathcal{C}$ as we discuss below. Moreover, the invariance of the partition function under the flop implies in particular the Gopakumar-Vafa invariants should not change during the transition, providing the parameters on both sides of the transition are matched appropriately; such a behaviour indeed follows in general from the topological vertex rules as shown in \cite{flop}, and the calculation below proves the consistency of our method with these results.
 
\begin{figure}[htb]
\begin{center}
\includegraphics[width=\textwidth]{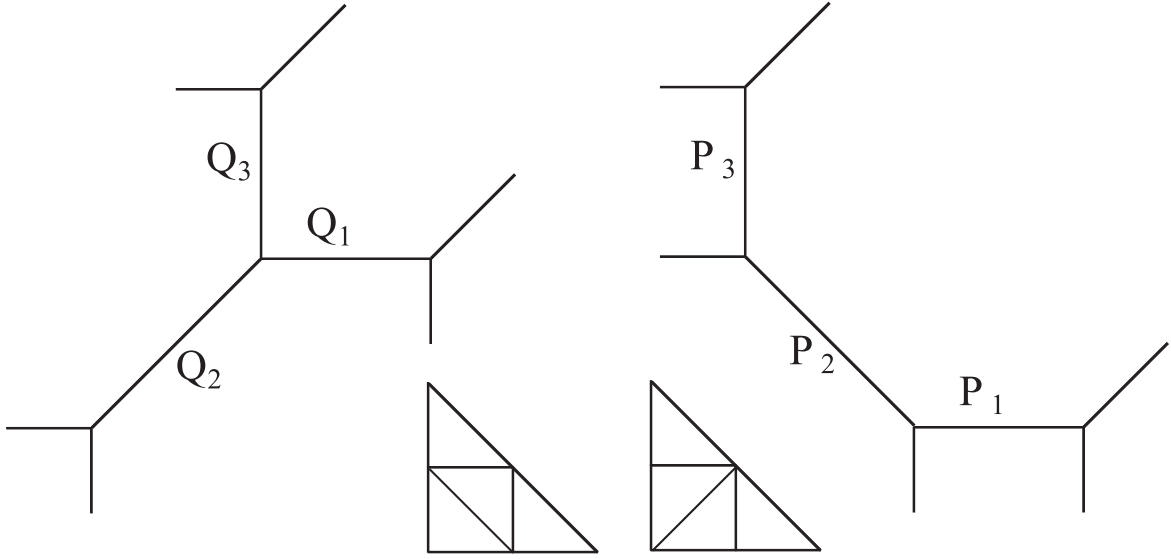}
\caption{The closed topological vertex $\mathcal{C}$ with its dual diagram (left) and the geometry after the flop $\mathcal{C}^{flop}$ (right).} \label{fig-flop-vertex}
\end{center}
\end{figure}

Let us focus on the closed topological vertex geometry, and the transition under which the conifold associated to $Q_2$ is flopped. We call the ensuing geometry $\mathcal{C}^{flop}$. The transition is presented in figure \ref{fig-flop-vertex}, and it is best understood in terms of a dual graph --- it is then represented by a tilt of a diagonal of a square corresponding to the conifold. The closed vertex on the left consists of three spheres meeting in one point, and after the flop it is replaced by a string of $\mathbb{P}^1$'s with two meeting points, and with a proper arrangement of bundles, as on the right side of the figure. We denote the parameters of the flopped geometry by $P_1, P_2, P_3$, and also express them in units of $g_s$ as
\begin{equation}
P_1 = q^{M_f}=e^{-s_1},\qquad P_2 = q^{L_f}=e^{-s_2}, \qquad P_3 = q^{N_f}=e^{-s_3}.  \label{P123-MLNf}
\end{equation}

How should the partition function for the flopped closed vertex look like after the transition? As mentioned above, it should be related to (\ref{closed-ver}) by a suitable identification of target space parameters. The geometry of the bundles in figure \ref{fig-flop-vertex} suggests the following relations
\begin{eqnarray}
Q_1 Q_2 & = &  P_1, \nonumber \\
Q_2  & = &  \frac{1}{P_2}, \label{t-s}  \\ 
Q_2 Q_3 & = &  P_3, \nonumber
\end{eqnarray}
from which it follows that $Q_1=P_1 P_2,\ Q_3=P_2 P_3,\ Q_1 Q_2 Q_3 =P_1 P_2 P_3, \ Q_1 Q_3 = P_1 P_2^2 P_3$. Substituting these into (\ref{closed-ver}) the only terms with negative powers we obtain are $P_2^{-n}$. Upon analytic continuation these should turn into $P_2^n$ together with appropriate change in the classical part of the free energy, as is the case for the conifold. Thus we expect the following instanton contribution to the partition function of $\mathcal{C}^{flop}$
\begin{equation}
Z^{\mathcal{C}^{flop}} = \exp\sum_{n>0}\frac{-P_1^n P_2^n - P_2^n - P_2^n P_3^n + P_1^n + P_1^n P_2^{2n} P_3^n + P_3^n - P_1^n P_2^n P_3^n}{n[n]^2}, \label{flop-ver}
\end{equation}
We first show by explicit calculation this is the correct result, and afterwords analyze classical contributions.

\begin{figure}[hbt]
\begin{center}
\includegraphics[width=0.4\textwidth]{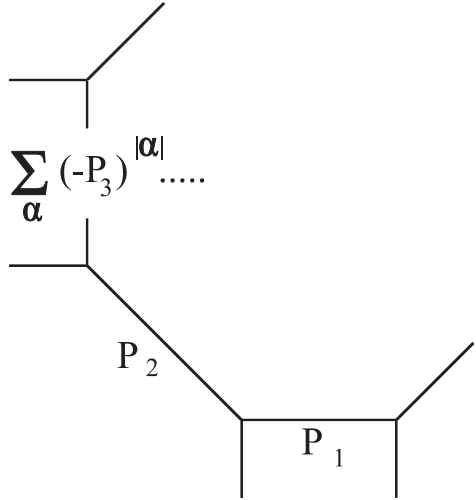}
\caption{The flopped closed topological vertex $\mathcal{C}^{flop}$ as a strip with attached $\mathbb{P}^1$ of the size determined by $P_3$.} \label{fig-flop-vertex-glue}
\end{center}
\end{figure}

To prove that (\ref{flop-ver}) is indeed correct we again use topological vertex rules. Similarly as for the closed vertex, we consider the flopped geometry as a strip with an additional $\mathbb{P}^1$ attached
$$
Z^{\mathcal{C}^{flop}} = \sum_{\alpha}  s_{\alpha^t}(q^{\rho}) (-P_3)^{|\alpha|} Z^{flop}_{\alpha} \, \Big[ (-1)^{|\alpha|} q^{-\kappa_{\alpha}/2} \Big],
$$
the factors in square brackets originating from a nontrivial framing of the additional sphere. This is presented in figure \ref{fig-flop-vertex-glue}, with $Z^{flop}_{\alpha}$ corresponding to the amplitude on the strip corresponding to a string of spheres $P_2-P_1$ 
\begin{eqnarray}
Z^{flop}_{\alpha} & = &  s_{\alpha}\, \{\alpha \bullet \}_{P_2} \{\alpha \bullet \}_{P_1 P_2} [\bullet \bullet ]_{P_1} = \nonumber \\
& = & e^{ \sum_n \frac{-P_2^n - P_1^n P_2^n + P_1^n}{n[n]^2} } \, s_{\alpha}(q^{\rho}) \, \prod_k (1-P_2 q^k)^{C_k(\alpha)} (1-P_1 P_2 q^k)^{C_k(\alpha)}, \nonumber
\end{eqnarray}
where we again used rules from the appendix \ref{app-strip} (now we read vertices from left to right, and they are of the types $A_{\alpha}-B-B$). Now we use (\ref{schur-hooks}) to write the full amplitude as
$$
Z^{\mathcal{C}^{flop}} = e^{ \sum_n \frac{-P_2^n - P_1^n P_2^n + P_1^n}{n[n]^2} } 
\sum_{\alpha} (q P_3)^{|\alpha|}  q^{2n(\alpha)}\prod_{(i,j)\in \alpha} \frac{1-P_2\, q^{j-i}}{1-q^{h(i,j)}}\, \frac{1-P_1 P_2\, q^{j-i}}{1-q^{h(i,j)}}.
$$
Finally, after identification (\ref{P123-MLNf}), using equality (\ref{schur-finite}) and summing over $\alpha$ we get
$$
Z^{\mathcal{C}^{flop}} = e^{ \sum_n \frac{-P_2^n - P_1^n P_2^n + P_1^n}{n[n]^2} } e^{ \sum_n \frac{P_3^n - P_2^n P_3^n - P_1^n P_2^n P_3^n + P_1^n P_2^{2n} P_3^n }{n[n]^2} }, 
$$
which is indeed the same as the expected result (\ref{flop-ver}) and consistent with \cite{flop}. 

\begin{figure}[htb]
\begin{center}
\includegraphics[width=0.8\textwidth]{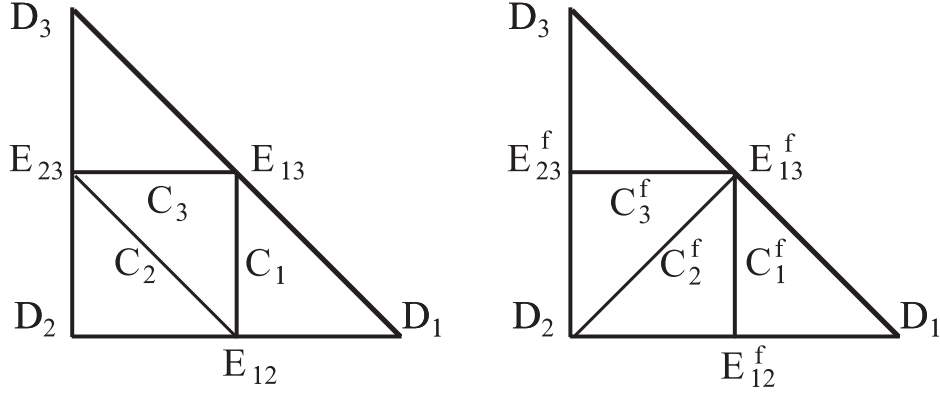}
\caption {Two geometries $\mathcal{C}$ and $\mathcal{C}^{flop}$ as a symmetric (left) and asymmetric (right) resolution of $\mathbb{C}^3/\mathbb{Z}_2 \times \mathbb{Z}_2$.} \label{fig-homologies}
\end{center}
\end{figure}

Let us finally turn to the issue of polynomial contributions. As mentioned above, those for $\mathcal{C}^{flop}$ and $\mathcal{C}$ are related to each other due to the invariance of the full partition function. We will show this relation is consistent with the values of cubic intersection numbers. These intersection numbers can be derived from the description of homology classes given in \cite{F-theory}, where both geometries are discussed as different resolutions of $\mathbb{C}^3/\mathbb{Z}_2\times\mathbb{Z}_2$ orbifold: the closed topological vertex is a symmetric resolution, and its flop an asymmetric resolution. The homology structure is in fact encoded in the dual diagram, with vertices corresponding to divisors and internal intervals to compact curves (each one arising as an intersection of two divisors at the end of the interval), as shown in figure \ref{fig-homologies}. There are three divisors $D_i$, $i=1,2,3$, in the singular orbifold, and additional three: $E_{ij}$ in its symmetric resolution, or $E_{ij}^{f}$ in asymmetric resolution, for $i\neq j$. The compact curves are the familiar by now three $\mathbb{P}^1$'s: $C_i$ with sizes given by $t_i$ for the closed vertex geometry and $C_i^f$ with sizes $s_i$ for its flop. For completness, let us recall the intersection numbers derived in \cite{F-theory}. For the closed topological vertex these are
\begin{center}
\begin{tabular}{c|c c c c c c}
   & $D_1$ & $D_2$ & $D_3$   & $E_{23}$ & $E_{13}$ & $E_{12}$ \\
\hline
$C_1 = E_{12}\cap E_{13}$ & $1$ & $0$ & $0$ & $1$ & $-1$ & $-1$ \\
$C_2 = E_{12}\cap E_{23}$ & $0$ & $1$ & $0$ & $-1$ & $1$ & $-1$ \\
$C_3 = E_{13}\cap E_{23}$ & $0$ & $0$ & $1$ & $-1$ & $-1$ & $1$
\end{tabular}
\end{center}
whereas for its flop
\begin{center}
\begin{tabular}{c|c c c c c c}
   & $D_1$ & $D_2$ & $D_3$   & $E^f_{23}$ & $E^f_{13}$ & $E^f_{12}$ \\
\hline
$C^f_1 = E^f_{12}\cap E^f_{13}$ & $1$ & $1$ & $0$ & $0$ & $0$ & $-2$ \\
$C^f_2 = E^f_{12}\cap E^f_{23}$ & $0$ & $-1$ & $0$ & $1$ & $-1$ & $1$ \\
$C^f_3 = E^f_{13}\cap E^f_{23}$ & $0$ & $1$ & $1$ & $-2$ & $0$ & $0$
\end{tabular}
\end{center}

From these intersection numbers we can deduce the form of genus zero prepotentials $F_0$ in terms of sizes of $\mathbb{P}^1$'s. In general
$$
F_0 = \frac{1}{6} J^3,
$$
with $J$ the Poincare dual to the K\"ahler form. For the closed vertex geometry $\mathcal{C}$ it can be parametrized by $e_i$ as
$$
J^{\mathcal{C}} = e_1 E_{12} + e_2 E_{13} + e_3 E_{23},
$$
so that the sizes of $\mathbb{P}^1$'s are given by
\begin{eqnarray}
t_1 & = & J^{\mathcal{C}} \cap C_1 = -e_1 -e_2 +e_3, \nonumber \\
t_2 & = & J^{\mathcal{C}} \cap C_2 = -e_1 +e_2 -e_3, \nonumber \\
t_3 & = & J^{\mathcal{C}} \cap C_3 =  e_1 -e_2 -e_3. \nonumber
\end{eqnarray}
Expressing $e_i$ in terms of $t_i$ and combining with genus one contributions (which depend on a single parameter $b$ due to symmetry in the geometry), the classical free energy (\ref{F-class}) for the closed topological vertex reads
\begin{equation}
F^{\mathcal{C}}_{class} = \Big(\frac{t_1^3 + t_2^3 + t_3^3}{6g_s^2} + \frac{t_1^2 t_2 + t_1 t_2^2 + t_1^2 t_3 + t_3 t_1^2 + t_2^2 t_3 + t_3 t_2^2}{4g_s^2} + \frac{t_1 t_2 t_3}{2g_s^2} \Big)  + \frac{b}{24}(t_1+t_2+t_3). \label{F-class-C}
\end{equation}
Similarly, for the flopped geometry the K\"ahler form has the Poincare dual
$$
J^{\mathcal{C}^{flop}} = e^f_1 E^f_{12} + e^f_2 E^f_{13} + e^f_3 E^f_{23},
$$
and the sizes of $\mathbb{P}^1$'s are
\begin{eqnarray}
s_1 & = & J^{\mathcal{C}^{flop}} \cap C^f_1 = -2e^f_1, \nonumber \\
s_2 & = & J^{\mathcal{C}^{flop}} \cap C^f_2 = e^f_1 -e^f_2 +e^f_3, \nonumber \\
s_3 & = & J^{\mathcal{C}^{flop}} \cap C^f_3 = -2e^f_3. \nonumber
\end{eqnarray} 
Including genus one contributions (for symmetry reasons now depending on two parameters $c,d$), we get classical free energy
\begin{eqnarray}
F^{\mathcal{C}^{flop}}_{class} & = & \Big(\frac{s_1^3 + 2s_2^3 + s_3^3}{6g_s^2} + \frac{2s_1^2 s_2 + 2s_1 s_2^2 + s_1^2 s_3 + s_3 s_1^2 + 2s_2^2 s_3 + 2s_3 s_2^2}{4g_s^2} + \frac{s_1 s_2 s_3}{2g_s^2} \Big)  + \nonumber \\
& + & \frac{c(s_1+s_3) + d s_2}{24} = \frac{t_1^3 + t_3^3}{6g_s^2} + \frac{t_1^2 t_2 + t_1 t_2^2 + t_1^2 t_3 + t_3 t_1^2 + t_2^2 t_3 + t_3 t_2^2}{4g_s^2} + \frac{t_1 t_2 t_3}{2g_s^2}  + \nonumber \\
& + & \frac{c(t_1+t_3)+t_2(2c-d)}{24}, \label{F-class-Cflop}
\end{eqnarray}
where we used the identification (\ref{t-s}) which equivalently reads $s_1=t_1+t_2,\ s_2=-t_2,\ s_3 = t_3+t_2$. Finally these classical terms can be combined with quantum ones $F^{\mathcal{C}}=\log Z^{\mathcal{C}}$ and $F^{\mathcal{C}^{flop}}=\log Z^{\mathcal{C}^{flop}}$ given in (\ref{closed-ver}) and (\ref{flop-ver}). Indeed, the full partition function is now explicitly seen to be invariant under the flop
$$
F^{\mathcal{C}}_{class}(t_1,t_2,t_3) + F^{\mathcal{C}}(t_1,t_2,t_3) = F^{\mathcal{C}^{flop}}_{class}(s_1,s_2,s_3) + F^{\mathcal{C}^{flop}}(s_1,s_2,s_3), 
$$
under the identification between $t_i$ and $s_i$ (\ref{t-s}), and providing $c=d=b$. There are two important remarks to be made. Firstly, quantum genus zero contributions for $\mathcal{C}^{flop}$, given by trilogarithm (\ref{F-conifold}), are continued to negative $t_2$ using $Li_3 e^{t_2} - Li_3 e^{-t_2} \sim t_2^3/6$ (where we keep only a cubic term, the other being ambiguous for topological string). This continuation is precisely the origin of the well known shift in classical intersection numbers under the flop (in our case this is seen explicitly in expressions (\ref{F-class-C}) and (\ref{F-class-Cflop})). Secondly, the condition $c=d=b$ which must be enforced is just a statement that the genus one classical part after the flop is entirely determined by the geometry before the flop.


\section{Discussion} \label{sec-discuss}

In this paper a class of Calabi-Yau crystals has been widened to include a  model which corresponds to a geometry of the closed topological vertex $\mathcal{C}$. The model is an extension - or rather a truncation - of other known crystal models to a finite cube. The model has been explicitly derived from topological string computations. In parallel the corresponding geometry $\mathcal{C}$ and its flop transition have been discussed, and Gopakumar-Vafa invariants and classical contributions to the free energy for $\mathcal{C}^{flop}$ determined.

These considerations suggest two obvious directions for further studies. Firstly, one might generalize the crystal interpretation of topological strings finding configurations relevant for arbitrary toric geometries. In particular there is a class of geometries considered in \cite{GW-curves}, which are an extension of the closed topological vertex by strings of spheres attached to any of its constituent $\mathbb{P}^1$'s. It requires essentially a new idea possibly related to gluing of finite crystals discussed here. One might also consider D-brane configurations, both within the closed topological vertex and in general geometries. It would be interesting to find their statistical and geometrical interpretation in the spirit of \cite{va-sa} and \cite{ani}.

The other task would be to simplify topological vertex rules for a wide class of geometries. These rules have already been simplified for geometries whose dual toric diagram is a triangulation of a strip \cite{strip}. As we discussed our results can be understood as 'off-strip' calculations, and they require using some special identities for Schur functions. It would be desirable to generalize these methods in a coherent fashion, in a way which would allow to turn arbitrary topological vertex expressions to a closed form with various sums automatically performed. Similarly, it would be interesting to include D-branes into such a general framework.

It is likely that these two lines of development might parallel each other. Apart from calculational advantages, one particular goal of such a program would be to find a proper formulation of the Gopakumar-Vafa invariants, at least in the context of toric manifolds. The integral structure of topological string amplitudes is still a mystery. Nonetheless, the structure of these amplitudes as obtained from crystal models is consistent with the Gopakumar-Vafa expansion. On the other hand, crystal models are known to be related to Donaldson-Thomas invariants which do have strong mathematical foundations. If a scope of Calabi-Yau crystals were extended to arbitrary (toric) geometries as suggested above, it would hopefully lead to a consistent formulation of the Gopakumar-Vafa invariants in terms of Donaldson-Thomas invariants.


\bigskip

\bigskip

\centerline{\Large{\bf Acknowledgments}}

\bigskip

This is my pleasure to thank Robbert Dijkgraaf for all the discussions, reading the manuscript and giving valuable comments. I am grateful to Paul Aspinwall and Frederik Denef for explanations on classical geometry, Lotte Hollands for the notes, Amir-Kian Kashani-Poor for explanation of his work, Jacek Pawe\l czyk for discussions and suggestions and Ani Sinkovics for crystal conversations. I also thank Amsterdam String Theory Group for great hospitality and stimulating atmosphere. This research was supported by the NWO Spinoza Grant and MNiSW grant N202-004-31/0060 for years 2006/2007.

\bigskip


\appendix

\section{Young diagrams and Schur functions} \label{app-schur}

Probably the best known source on Schur functions is \cite{macdonald}. The issues relevant to topological string calculations have been presented in \cite{ps}. In this appendix we briefly recall the notation we use (which is the same as in \cite{ps}), and present the most important properties we need, together with some new identities which are crucial for present calculations.

Young diagrams or partitions are denoted by letters $P,R$ or Greek ones
$\alpha$ etc., and $^{t}$ means a transposition. For a partition
$R=(R_1,R_2,\ldots)$ with lengths of rows given by $R_i$, one defines
\begin{eqnarray}
|R| & = &\sum_{i} R_i, \nonumber\\
n(R) & = & \sum_{i} (i-1)R_i,\nonumber\\ 
\kappa_R & = & |R|+\sum_i R_i(R_i -2i) = 2\sum_{(i,j)\in R} (j-i), \nonumber 
\end{eqnarray}
where $(i,j)$ is a position of a certain box in the diagram. For such a single box,
its hook length is defined as
\begin{equation}
h(i,j)=R_i+R_{j}^{t}-i-j+1. \label{hook}
\end{equation}

Schur function for a partition $R$ is denoted as $s_R$. By $q^{R+\rho}$ we understand a string such that $x_i=q^{R_i-i+1/2}$ for $i=1,2,\ldots$. Thus
$$
s_R(q^{R+\rho})=s_R(q^{R_1-1/2},q^{R_2-3/2},\ldots).
$$
In particular
$$
s_R(q^{\rho})=s_R(q^{-1/2},q^{-3/2},\ldots).
$$
The following identity holds for Schur function with finite number of arguments
\begin{equation}
s_R(1,q,q^2,\ldots,q^{K-1}) = q^{n(R)} \prod_{(i,j)\in R} \frac{1-q^{K+j-i}}{1-q^{h(i,j)}}, \label{schur-finite}
\end{equation}
and for $K\to\infty$ this reduces to
\begin{equation}
s_R(q^{-\rho}) = q^{|R|/2+n(R)} \prod_{(i,j)\in R} \frac{1}{1-q^{h(i,j)}}.  \label{schur-hooks}
\end{equation}

A sum of Schur functions can be written as follows
\begin{eqnarray}
\sum_{P} s_{P}(x) s_{P^t}(y) & = & \prod_{i,j} (1+x_i y_j) = \label{schur-sum} \\
& = & \exp\Big[-\sum_{n,i,j} \frac{(-1)^n}{n} x_i^n y_j^n \Big]. \label{schur-sum-exp}
\end{eqnarray}
As noticed in \cite{strip}, with $x=q^{R+\rho}$ and $y=q^{R'+\rho}$ this allows to write the sum as
\begin{equation}
\sum_{P} s_{P}(q^{R+\rho}) s_{P^t}(-Q q^{R'+\rho}) = \exp\Big[-\sum_n \frac{Q^n}{n[n]^2} \Big] \, \prod_k (1-Qq^k)^{C_k(R,R')}, \label{sum-schur-RRbis}
\end{equation}
where coefficients $C_k(R,R')$ are given by
\begin{equation}
\sum_{i,j} q^{R_i-i+1/2} q^{R'_j-j+1/2} = \sum_k C_k(R,R') q^k + \frac{q}{(1-q)^2}.
\end{equation}
From this statement the following properties are more or less easily deduced
\begin{eqnarray}
C_k(R,R') & = & C_k(R',R), \nonumber \\
\sum_k C_k(R,R')q^k & = & \sum_k C_k(R^t,R'^t)q^{-k}, \nonumber \\
\sum C_k(R,R') & = & |R|+|R'|, \nonumber \\
\sum k C_k(R,R') & = & \frac{\kappa_R + \kappa_{R'}}{2}. \nonumber \\
\end{eqnarray}
In particular, $C_k(R)=C_k(R,\bullet)$ counts the number of boxes $(i,j)\in R$ with fixed $k=j-i$, and so $\sum_k C_k(R)=|R|$ and $2\sum_k kC_k(R)= \kappa_R$.


\section{Topological vertex on a strip} \label{app-strip}

There is a class of toric Calabi-Yau geometries whose dual diagrams are represented as a triangulation of a rectangle (or a \emph{strip}). Computation of their partition functions via topological vertex methods has been vastly simplified in \cite{strip}. As this simplification is quite convenient for a part of the calculations we perform, we briefly recall the \emph{rules on the strip}. We encourage a reader to consult \cite{strip} for more through presentation and proofs of these rules.

A diagram which can be drawn on a strip is a string of $\mathbb{P}^1$'s with parameters $Q_i$, each represented by an interval between vertices $i$ and $i+1$. For every such interval we introduce representations $R_i$ which we sum over according to topological vertex rules. Additionally, with every external leg of each vertex we can associate one fixed representation $\alpha_i$ (for the first vertex in a string there are two external legs, but one of them must be associated with a trivial representation $\bullet$; the same statement must hold for the last vertex).

The partition function for such a system can be expressed in terms of the quantities
\begin{equation}
\{\alpha_i \alpha_j\}_{Q_{ij}} =  \exp \Big[-\sum_n \frac{Q^n_{ij}}{n[n]^2} \Big] \prod_k (1-Q_{ij}q^k)^{C_k(\alpha_i,\alpha_j)}, \label{amir-bracket}
\end{equation}
$$
[\alpha_i \alpha_{j}]_{Q_{ij}} = \{ \alpha_i \alpha_j \}^{-1}_{Q_{ij}},
$$
according to the following rules:
\begin{itemize}
\item determine the type of the first ($i=1$) vertex to be A or B if respectively its topological vertex factor is given by $C_{\bullet\alpha_1 R_1}$ or $C_{\bullet R_1 \alpha_1}$

\item determine recursively the type A or B of all other vertices: $(j+1)$'th vertex is of the same type as $j$'th if the local bundle of the sphere $Q_j$ is $\mathcal{O}(-2)\oplus\mathcal{O}$, and of a different type if this bundle is $\mathcal{O}(-1)\oplus\mathcal{O}(-1)$

\item for each pair of vertices $(i,j)$ in a diagram (with $i<j$) introduce a suitable factor according to their types $(A/B,A/B)$:
\begin{eqnarray}
(A,B) & \to & \{\alpha_i \alpha_j\} _{Q_{ij}}, \nonumber \\
(B,A) & \to & \{\alpha_i^t \alpha_j^t\} _{Q_{ij}}, \nonumber \\
(A,A) & \to & [\alpha_i \alpha_j^t]_{Q_{ij}}, \nonumber \\
(B,B) & \to & [\alpha_i^t \alpha_j]_{Q_{ij}}, \nonumber
\end{eqnarray}
where $Q_{ij} = Q_i Q_{i+1} \cdots Q_{j-1}$.

\item the full amplitude, with external representations $\alpha_i$ fixed, is given by a product of all factors above together with Schur functions for all external representations
$$
Z_{\prod \alpha_i} = \prod_i s_{\alpha_i}(q^{\rho})\ \prod_{i<j} [\{\alpha_i^{\dag} \alpha_j^{\dag} \}]_{Q_{ij}},
$$ 
with appropriate choice of the pairing $[\{\,\}]$, and with or without transposition $\dag=t,\cdot$.

\end{itemize} 

For example, in figure \ref{fig-closed-ver-glue} the strip-part consists of two intervals $Q_1$ and $Q_2$. We denote the right-most vertex by $i=1$, the middle one with external representation $\alpha$ by $i=2$ and the left-most by $i=3$. There is no external representation on the first vertex, so we are free to choose its type to be e.g. A; then we recursively determine types of all three vertices to be $A-B_{\alpha}-A$ (for convenience we explicitly write external representations associated with vertices, if they are nontrivial). Then according to the rules above the amplitude reads 
\begin{equation}
Z_{\alpha} = s_{\alpha}\, \{\bullet \alpha \}_{Q_1} [\bullet \bullet ]_{Q_1 Q_2} \{\alpha^t \bullet \}_{Q_2}. \label{Zalpha}
\end{equation}




\end{document}